\documentstyle[epsf]{elsart}

\begin{document}
\begin{frontmatter}
\title{Manifestation of the magnetic depopulation
       of one--dimensional subbands in the optical
       absorption of acoustic magnetoplasmons in
       side--gated quantum wires}
\author[Hamburg]{C. Steinebach},
\author[Hamburg]{T. Kurth},
\author[Hamburg]{D. Heitmann}, and
\author[Reykjavik]{V. Gudmundsson}
\address[Hamburg]{Institut f{\"u}r Angewandte Physik und 
Zentrum f{\"u}r Mikrostrukturforschung, Universit{\"a}t
Hamburg, Jungiusstra{\ss}e 11, D-20355 Hamburg, Germany}
\address[Reykjavik]{Science Institute, University of Iceland, Dunhaga 3, 
IS-107 Reykjavik, Iceland}

\begin{abstract}
We have investigated experimentally and theoretically the far--infrared
(FIR) absorption of gated, deep--mesa--etched GaAs/Al$_x$Ga$_{1-x}$As 
quantum wires. To overcome
Kohn's theorem we have in particular prepared double--layered wires and 
studied the acoustic magnetoplasmon branch. We find oscillations in the 
mag\-net\-ic--field dispersion of the acoustic plasmon which are traced back 
to the self--consistently screened density profile in its dependence 
on the magnetic depopulation of the one--dimensional subbands.
\end{abstract}
\thanks{This work was supported by the Deutsche Forschungsgemeinschaft 
through SFB508, Grant He1938/6-1+6-2, the Graduiertenkolleg 
``Physik Nanostrukturierter Festk\"orper'',
and the Icelandic Science Foundation.}
\end{frontmatter}

Far--infrared  spectroscopy has proven to be a powerfull tool
to study the elementary excitations of low--dimensional electron
systems such as quantum wires and quantum dots 
\cite{Demel88:12732,Drexler92:12849,Demel91:2657,Bollweg96:2774}. 
It is well known that the FIR--absorption in parabolically confined
quantum wires and quantum dots is governed by Kohn's theorem
\cite{Brey89:10647,Maksym90:108,Kohn61:1242}.
A considerable effort was made to overcome the restrictions of 
this theorem which does not allow the observation of the
internal structure of the systems. 
In this paper we concentrate our attention on the acoustic magnetoplasmon
mode in double--layered quantum wires. Acoustic modes,
which correspond to an out--of--phase motion of the
electrons in the two sheets, are not restricted by Kohn's
theorem and allow a direct observation of effects of the 
electron--electron interaction
\cite{Demel88:12732,DasSarma81:805,Fasol86:2517,DasSarma91:11768}.\\
In earlier measurements of acoustic magnetoplasmons in double--layered
quantum wires \cite{Demel88:12732} the distance $D$ between the wires
was about 300 nm. In order
to increase the Coulomb coupling between the wires, we have prepared 
wires on the basis of two symmetrically grown, modulation--doped 7 nm 
quantum wells, separated by a 55 nm doped barrier. After the deposition
of 15 nm SiO$_x$N$_y$ by PECVD a NiCr gate has
been prepared on top and on the sidewalls of the deep--mesa etched
wires. These sidegates allow an effective tuning of the 1D subband structure
and the 1D carrier density. The samples have been studied by 
Fourier--transform transmission spectroscopy. A magnetic field is
applied in the growth direction perpendicular to the wires.
The magnetic--field dispersion
of the observed peaks is displayed in  Fig.~\ref{figure1}. The carrier
density is $n_{1D}=1\times 10 ^{7}$ cm$^{-1}$ per single wire. 
We observe two modes, the upper mode $\omega_{o,1}$ is an optical mode
which carries most of the oscillator strength. The low--energy mode
$\omega_{a,1}$ is an acoustic mode. Besides we observe also higher modes 
which are labeled as $\omega_{o,3}$ and $\omega_{a,3}$. 
The optical branch satisfies Kohn's theorem except for a wiggle in 
the vicinity of $E=2\omega_c$. This is a precursor of the well
known Bernstein splitting \cite{Bernstein58:10,Gudmundsson95:17744}.
What is new are the oscillations in the acoustic branch labeled by
$\nu=6$ and $\nu=4$. From measurements of the magneto resistance
we find that the observed oscillations reflect directly the magnetic
depopulation of 1D subbands. 

\epsfxsize=7.0cm
\hspace*{3cm}{\centering {\epsfbox{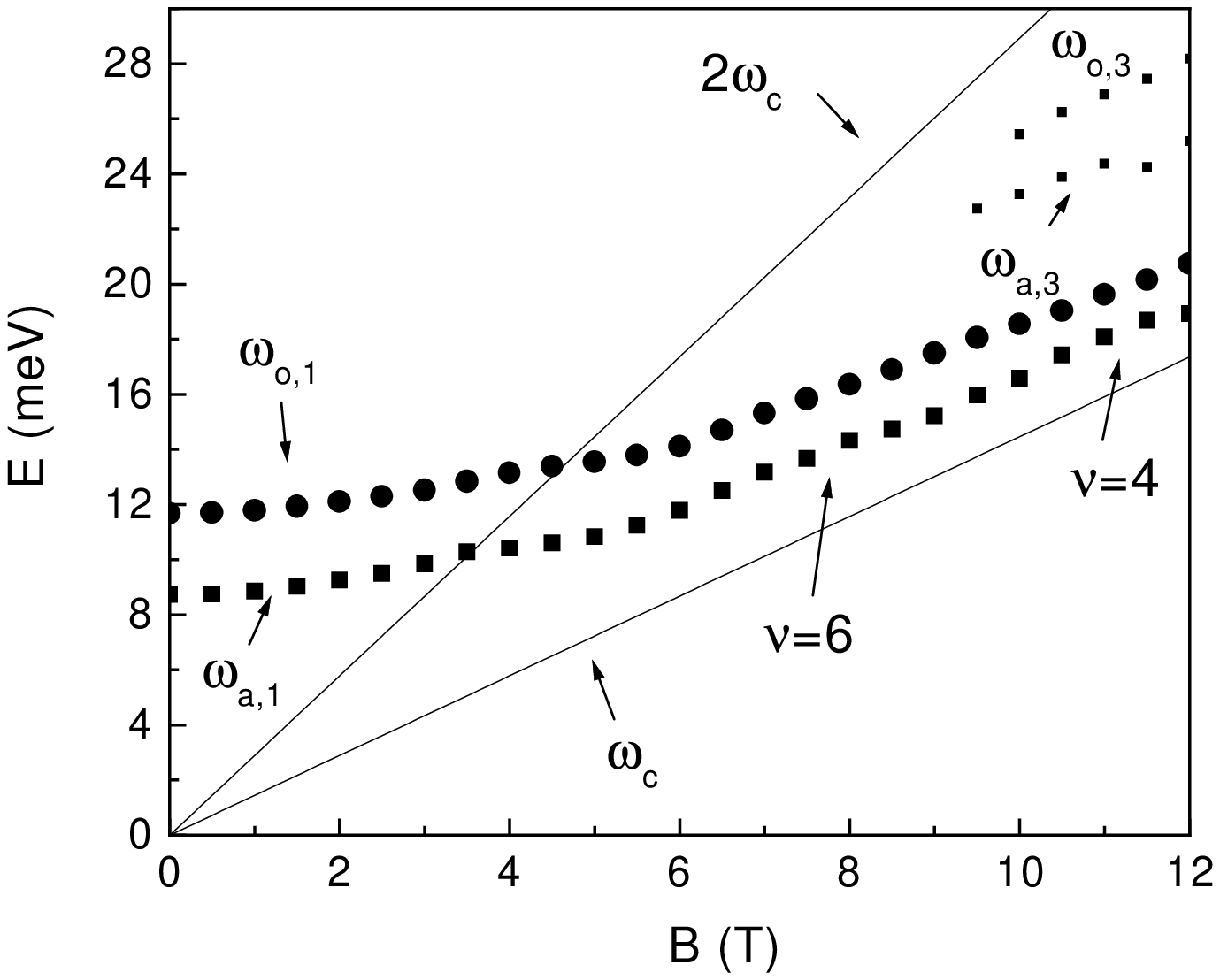}}}
\begin{figure}[tbh]
\caption{Magnetic--field dispersion of the measured optical and acoustic
modes in a double--layered system of quantum wires. Magnetic fields which 
correspond to even filling factors are marked by $\nu=6$ and $\nu=4$. The gate
voltage is $V_G=0.4$ V which gives a linear density of 
$n_{1D}=1 \times 10^7$ cm$^{-1}$ per single wire.} 
\label{figure1}
\end{figure}

\noindent In the following we want to investigate the  origin of 
these oscillations.
We have therefore performed self--consistent Hartree calculations of the
subband structure of double--layered quantum wires. The absoption spectrum
is calculated within the random--phase approximation (RPA) 
\cite{DasSarma91:11768}. For the 
 calculation we assume that the confinement in growth direction $z$
is strong enough to neglect the finite thickness of the electron system. 
The electrons are located in either one of the wires. 
They are coupled by direct Coulomb interaction only.
In the experiment 12 subbands are occupied at $B=0$ T. The layer separation
is $D=55$ nm where tunneling is unimportant.\\
  To facilitate 
the numerical calculation we consider a smaller system with only 3
occupied subbands and smaller $D=10$ nm. We thus attain a good qualitative 
agreement with experiment rather than quantitative.  
In order to observe acoustic modes an asymmetry between the two sheets
has to be introduced. This can be done either by different external 
potentials or electron densities in the wires \cite{stei97}. Another possibilty
is to apply an exciting potential that has a spatial modulation in  growth
direction, (caused by the grating coupler on top of the sample).  
For our calculations we consider two identical parabolic wires and an
external field with a nonzero component in growth direction, guaranteeing
that our results do not depend on the special choice of the confining 
potentials. 
Figure~\ref{figure2} shows the calculated dispersion for the
case of identical parabolic external confinements of $\hbar \omega_0=8$ meV.\\
\epsfxsize=7.5cm
\hspace*{3.5cm}{\centering{\epsfbox{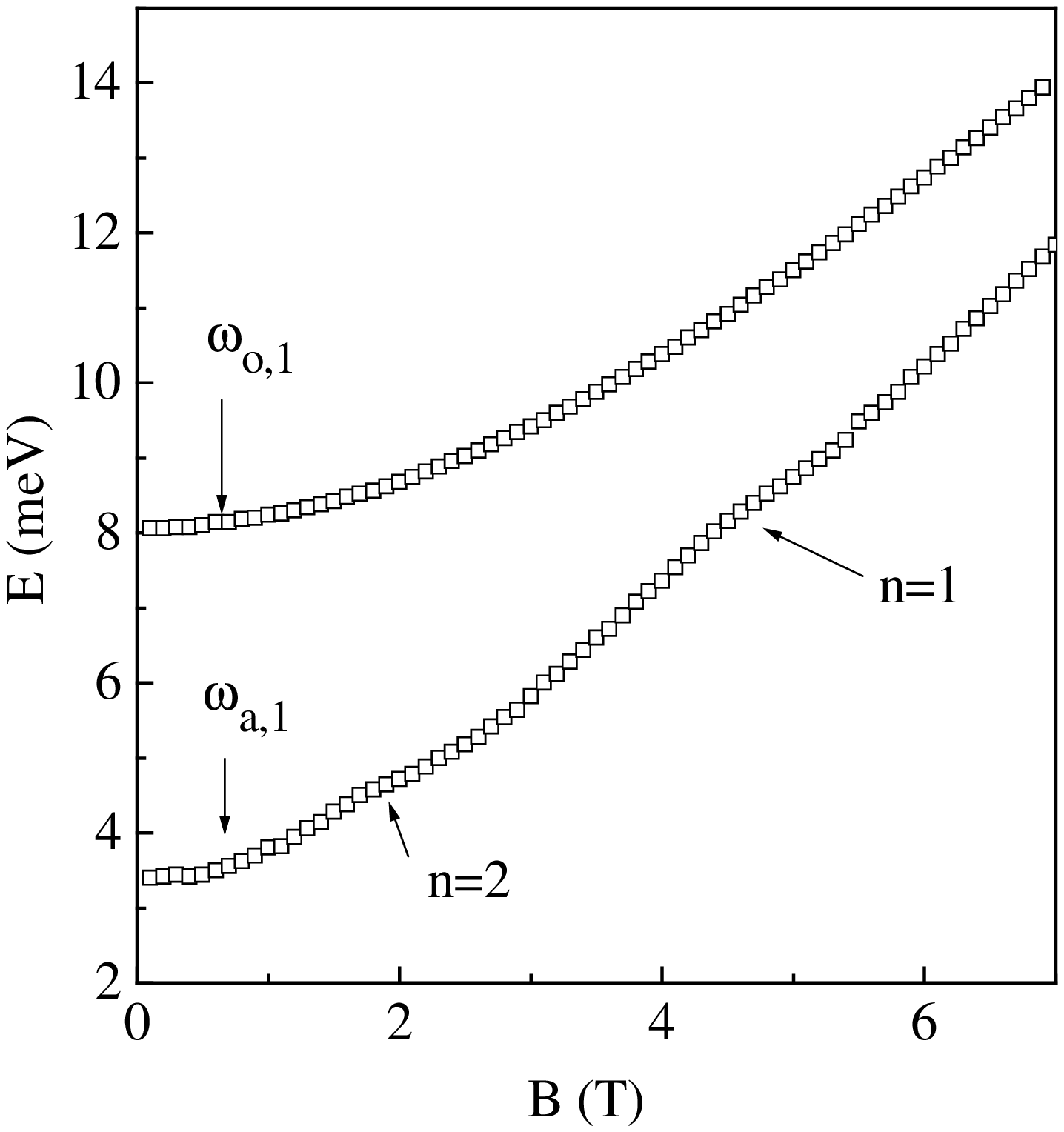}}}
\begin{figure}[tbh]
\caption{Calculated magnetoplasmons for an external potential of 
$\hbar \omega_0=8$ meV. The linear carrier density is $n_{1D}=2.5\times
10^6$ cm$^{-1}$ per wire. The interlayer distance is $D=10$ nm.} 
\label{figure2}
\end{figure}

\noindent The magnetic fields where the chemical potential $\mu $ is just below
the second and the third spin--degenerate subband are labeled by 
$n=1$ and $n=2$, respectively. At these magnetic fields we find
local maxima in the acoustic branch labeled by $\omega_{a,1}$. Since 
the confining potential is parabolic the optical branch follows the
dispersion law $\omega_{o,1}^2=\omega_0^2+\omega_c^2$, where $\omega_c=eB/m^*$
is the cyclotron frequency. The oscillations
in the acoustic branch  directly reflect the screening properties
and the resulting density profile \cite{Bollweg96:2774}
of the electron gas in the ground state which depend strongly
on where the chemical potential is located with respect to the
subband edges. In the classical limit of broad wires with high
electron density the screening results in compressible and 
incompressible stripes \cite{chkl92:4026}.
This is demonstrated in Fig.~\ref{figure3} where the
self--consistent Hartree potential is plotted. At $B=2.5$ T where
we have a local minimum in the resonance position the wire develops 
a flat region in its center. This is due to the good screening 
of the electron gas when the chemical potential $\mu$ is just above
the edge of the second subband. The flat region coincides with a small
subband quantization and gives rise to the minimum. For a field of
$B=4.5$ T the second subband is just unoccupied which results in
a low density of states around $\mu $ and weak screening properties.
The flat region in the center of the wire has vanished and the subband 
quantization due the Hartree potential is larger than in the case
of $B=2.5$ T. These results for the field--dependent screening 
agree well with earlier calculations of Suzuki and Ando \cite{Suzu93:2986}.\\
\epsfxsize=7.5cm
\hspace*{3.25cm}\epsfbox{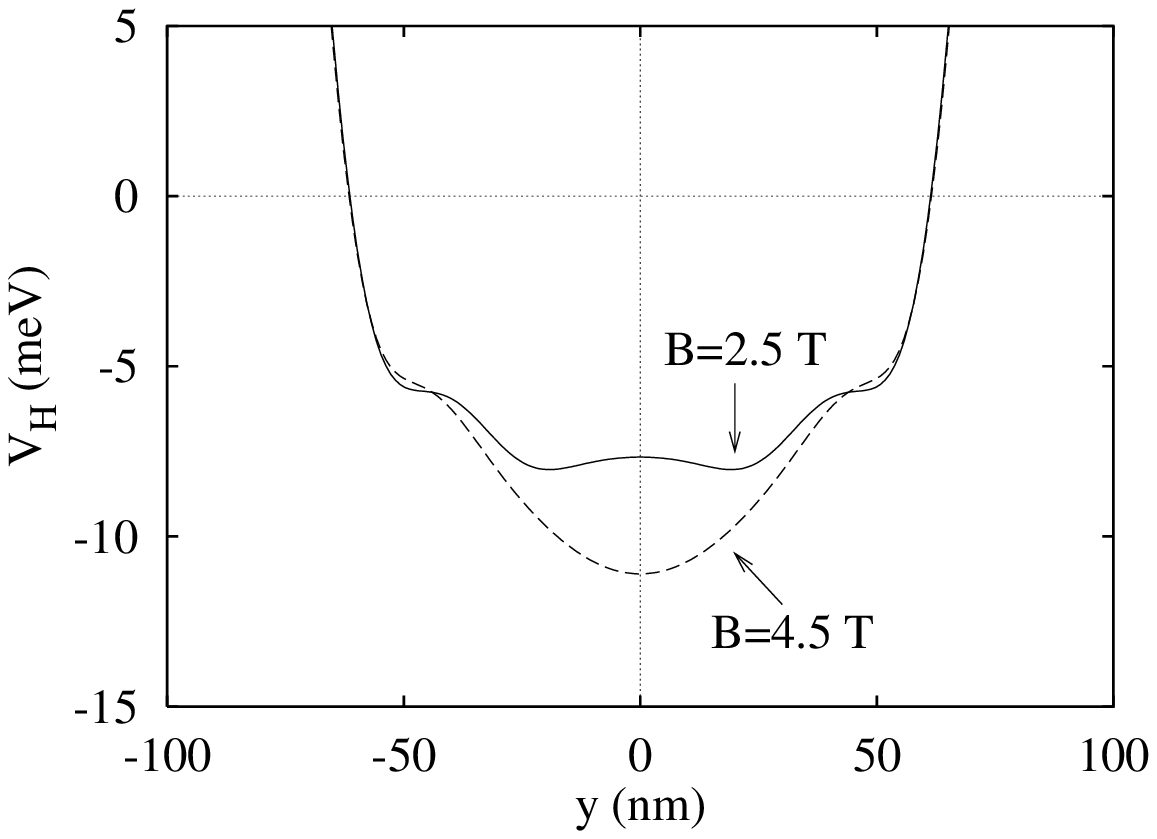}
\begin{figure}[tbh]
\caption{Self--consistent Hartree potential for the magnetic fields $B=2.5$ T
(solid line) and $B=4.5$ T (dashed line). The chemical potential is $\mu=0$.} 
\label{figure3}
\end{figure}

\noindent In summary we have demonstrated experimentally and theoretically that
acoustic modes provide a powerfull method to study the internal structure of
quantum wires by means of optical measurements.

\bibliographystyle{prsty}

\end{document}